\documentstyle [12pt] {article}
\title{SIMPLE SOLUTIONS OF THE PROTON SPIN CRISIS AND SUPERSYMMETRY CRISIS}

\author{Vladan Pankovi\'c,\\
Department of Physics, Faculty of Sciences, 21000 Novi Sad,\\ Trg
Dositeja Obradovi\'ca 4, Serbia, \\vladan.pankovic@df.uns.ac.rs}

\date {}
\begin {document}
\maketitle \vspace {0.5cm} PACS number: 14.20Dh, 14.40.$\pm n $,
12.60. Jv

\begin {abstract}
In this work we suggest a simple theoretical model of the proton
able to effectively solve proton spin crisis. Within domain of
applicability of this simple model proton consists only of two u
quarks and one d quarks (two of which have spin opposite to proton
and one identical to proton) and one neutral vector phi meson
(with spin two times larger than proton spin and directed
identically to proton spin). This model is in full agreement not
only with existing DIS experiments, but also with spin and
electric charge conservation as well as in a satisfactory
agreement with rest mass-energy conservation (since phi meson mass
is close to proton rest mass). Our model opens an interesting
possibility of the solution of the quarks and leptons families
problem (proton is not an absolutely non-strange particle, but
only a particle with almost totally effectively hidden strange).
Also we suggest a possible "first step" toward the solution of the
supersymmetry crisis using so-called superexclusion principle.
According to this principle usual particles (electron, neutrino, …
) can exist actually and virtually, while their supersymmetric
partners, s-particles (selectron, neutralino, …) can exist
(super)exclusively virtually but not actually.
\end {abstract}

 In this work we shall suggest a simple theoretical model of the
 proton able to effectively solve proton spin crisis. Within domain
 of applicability of this simple model proton $p$ consists only of
 two $u$ quarks and one $d$ quarks (two of which have spin opposite
 to proton and one identical to proton) and one neutral vector phi
 meson $\phi$ (with spin two times larger than proton spin and
 directed identically to proton spin). This model is in full agreement
 not only with existing deep inelastic scattering, DIS, experiments,
 but also with spin and electric charge conservation as well as in
 a satisfactory agreement with rest mass-energy conservation (since
 phi meson mass is close to proton rest mass). Our model opens an
 interesting possibility of the solution of the quarks and leptons
 families problem (proton is not an absolutely non-strange particle,
 but only a particle with almost totally effectively hidden strange).
 We paraphrase, "all lucky families are lucky in the unique way".

Also we suggest a simple "first step" toward the solution of the
supersymmetry crisis using so-called superexclusion principle.
According to this principle usual particles (electron, neutrino, …
) can exist actually and virtually, while their supersymmetric
partners, s-particles (selectron, neutralino, …) can exist
(super)exclusively virtually but not actually.

Consider a proton, p, in such simplified way, in which it consists
of only three (interacting) sub-systems, two up quarks, $u$ and
$u$, and one down quark $d$ (gluons and other proton constituents
will not be considered here explicitly).

Further, suppose that on so-described $p$ only experiment of deep
inelastic scattering DIS (with muons and similar systems) [1]-[4]
can be made. This experiment can be considered as the sub-systemic
measurement of spin at one (of the three possible proton quarks).

In mentioned experiment proton is initially (before scattering,
i.e. measurement) prepared in eigen spin state in appropriate
direction $|\frac {1}{2}>$ with corresponding spin value $\frac
{1}{2}$ (in the reduced Planck constant units).

According to introduced suppositions we can theoretically
considered the following proton initial (before measurement) spin
state
\begin {equation}
   |p 3> = \frac {1}{3^{\frac {1}{2}}} |\frac {1}{2}>_{1}|\frac {1}{2}>_{2}|-\frac {1}{2}>_{3}
   + \frac {1}{3^{\frac {1}{2}}} |\frac {1}{2}>_{1}|-\frac {1}{2}>_{2}|\frac {1}{2}>_{3}
   +  \frac {1}{3^{\frac {1}{2}}}|-\frac {1}{2}>_{1}|\frac {1}{2}>_{2}|\frac {1}{2}>_{3}
\end {equation}
where $|\pm \frac {1}{2}>_{j}$ can be considered as the eigen
state of spin in the same direction of the proton spin with
corresponding eigen values $\pm \frac {1}{2}$ for subsystem $j=1,
2, 3$ . Subsystem $j =1, 2, 3$ , can be, in principle, any of two
u quarks or d quark

Finally, suppose that mentioned DIS subsystemic spin measurement
will be always realized at the 1 subsystem. Since state (1) is
symmetric in respect to permutation of the index $j$ values
introduced supposition does not mean any restriction on the real
DIS experiments.

Suppose now that at proton spin state (1), instead of mentioned
subsystemic spin measurement on 1 only, subsystemic measurements
of spin at all 1, 2 and 3 are realized simultaneously. Then, after
these measurements, with the same probabilities $\frac {1}{3}$,
proton spin state (1) turns out either in $|\frac
{1}{2}>_{1}|\frac {1}{2}>_{2}|-\frac {1}{2}>_{3}$ or in $|\frac
{1}{2}>_{1}|-\frac {1}{2}>_{2}|\frac {1}{2}>_{3}$ or in $|-\frac
{1}{2}>_{1}|\frac {1}{2}>_{2}|\frac {1}{2}>_{3}$. In any of these
states proton spin has the same value $\frac {1}{2}$ on the one
hand, and, on the other hand two quarks have the same spin as the
proton while one quark has spin opposite to proton.

But in DIS there is principally different situation (discussion of
this subject but without explicit mathematical formalism is given
in [5]). Before mentioned DIS subsystemic measurement proton spin
state (1) must be transformed in the following expression
\begin {equation}
   |p 3> = (\frac {2}{3})^{\frac {1}{2}} |\frac {1}{2}>_{1}[\frac {1}{2^{\frac {1}{2}}} (|\frac {1}{2}>_{2}|-\frac {1}{2}>_{3} + |-\frac {1}{2}>_{2}|\frac {1}{2}>_{3})] + (\frac {1}{3})^{\frac {1}{2}} |-\frac {1}{2}>_{1}|\frac {1}{2}>_{2}|\frac {1}{2}>_{3}.
\end {equation}
It implies that after DIS measurement, with probability $\frac
{2}{3}\simeq 0.66$, measured quark, 1, are described by spin eigen
state $|\frac {1}{2}>_{1}$ with spin $\frac {1}{2}$ equivalent to
proton spin. Simultaneously two other quarks, 2 and 3, do a
complex system $2+3$ (that does not admit any separation in 2 and
3) described by entangled quantum state
$[\frac{1}{2^{\frac{1}{2}}} (|\frac {1}{2}>_{2}|-\frac {1}{2}>_{3}
+ |-\frac {1}{2}>_{2}|\frac {1}{2}>_{3})]$ with zero spin even if
neither 2 nor 3 is described by any spin eigen state. On the other
hand, after DIS measurement, with probability $\frac {1}{3}\simeq
0.33$, measured quark, 1, is described by spin eigen state
$|-\frac {1}{2}>_{1}$ with spin $-\frac {1}{2}$ opposite to proton
spin. Simultaneously, two other quarks, 2 and 3, are described by
spin eigen states $|\frac {1}{2}>_{2}$  and $|\frac {1}{2}>_{3}$
with spins $\frac {1}{2}$ and $\frac {1}{2}$  both equivalent to
proton spin.

All this represents an interesting result. Namely in real DIS
experiments it is obtained that with probability about 0.66
measured quark has spin $-\frac {1}{2}$ opposite to proton spin,
while with probability about 0.33 measured quark has spin $\frac
{1}{2}$  equivalent to proton spin, what is in some way
"reciprocal" to previous theoretical predictions based on (2).

Now we shall suggest the simplest generalization of the mentioned
theoretical description of the proton spin state which would
satisfy real DIS experiments.

Suppose now that proton, more accurately speaking, consists of
four subsystems, three previously mentioned quarks and an
additional forth subsystem 4 whose characteristics will be
determined later.

According to introduced supposition we can theoretically
considered the following proton spin state before measurement
\begin {equation}
   |p 4> = \frac {1}{3^{\frac {1}{2}}} |-\frac {1}{2}>_{1}|-\frac {1}{2}>_{2}|\frac {1}{2}>_{3}|1>_{4}+
   \frac {1}{3^{\frac {1}{2}}} |-\frac {1}{2}>_{1}|\frac {1}{2}>_{2}|-\frac {1}{2}>_{3}|1>_{4}
\end {equation}
   \[+ \frac {1}{3^{\frac {1}{2}}}|\frac {1}{2}>_{1}|-\frac {1}{2}>_{2}|-\frac
   {1}{2}>_{3}|1>_{4}\]
  \[ = (\frac {1}{3^{\frac {1}{2}}} |-\frac {1}{2}>_{1}|-\frac {1}{2}>_{2}|\frac {1}{2}>_{3}
   + \frac {1}{3^{\frac {1}{2}}} |-\frac {1}{2}>_{1}|\frac {1}{2}>_{2}|-\frac
   {1}{2}>_{3}\]
   \[+ \frac {1}{3^{\frac {1}{2}}}|\frac {1}{2}>_{1}|-\frac {1}{2}>_{2}|-\frac {1}{2}>_{3})
   |1>_{4}\]

where $|1>_{4}$ can be considered as the eigen state of the spin
of subsystem 4 in the same direction of the proton spin with
corresponding spin eigen value 1. Then, obviously, total spin
corresponding to any superposition term equals $2(-\frac
{1}{2})+\frac {1}{2}+1=\frac {1}{2}$ so that it is equal to the
proton spin. In this way there is spin conservation.

Further, suppose quite naturally that electrical charge is
conserved too. It implies that 4 is electrically neutral, i.e.
that its electrical charge equals 0. Then, obviously, charge of
the proton is simply sum of the charges of two $u$ and one $d$
quark, i.e.  $2 (\frac {2}{3})-\frac {1}{3} = 1$ (in elementary
electrical charge $e$ unit system).

Now, since it is supposed that according to spin and electrical
charge conservation law subsystem 4 has spin 1 and 0 electrical
charge it is not hard to see that such 4 subsystem would be a
neutral vector meson. It can be supposed that this meson is $\phi$
meson since its rest mass that equals about 1 020 MeV closest (in
respect to mass of all other neutral vector mesons) to proton mass
that equals about 938 MeV.

As it is well known common rest mass of two u and one d quark
equals only about 11 MeV what is about 1 percent of the proton
rest mass. For this reason it is usually supposed that proton mass
is dominantly result of the dynamical interaction between quarks
via gluons even if such prediction is very hard for exact
calculations. (It implies too that proton spin crisis must be
solved via introduction of the angular momentum of the quarks and
gluons what is also very hard experimentally verified). However,
if it is supposed that proton effectively, i.e. in a satisfactory
approximation, consists of two up, one down quark and phi meson,
then it represents an almost "static" supersystem.

Thus, in our model of the proton, before mentioned DIS subsystemic
measurement proton spin state (3) must be transformed in the
following expression
\begin {equation}
   |p 4> = (\frac {2}{3})^{\frac {1}{2}} [|-\frac {1}{2}>_{1}[\frac {1}{2^{\frac {1}{2}}} (|-\frac {1}{2}>_{2}|\frac {1}{2}>_{3} + |\frac {1}{2}>_{2}|-\frac {1}{2}>_{3})]|1>_{4} +
                    \frac {1}{3^{\frac {1}{2}}} |\frac {1}{2}>_{1}|-\frac {1}{2}>_{2}|-\frac {1}{2}>_{3}|1>_{4}    .
\end {equation}

It is very important to be pointed out that subsystem 4, i.e. phi
meson, in (4) is non-entangled with any of quarks 1, 2 or 3. on
the other hand since 4 is electrically neutral it is practically
unobservable for DIS experiment.

All this implies that after DIS measurement, with probability
$\frac {2}{3}\simeq 0.66$, measured quark, 1, is described by spin
eigen state $|-\frac {1}{2}>_{1}$ with spin $-\frac {1}{2}$
opposite to proton spin. Simultaneously two other quarks, 2 and 3,
do a complex system 2+3 (that does not admit any separation in 2
and 3) described by entangled quantum state $[\frac {1}{2^{\frac
{1}{2}}} (|-\frac {1}{2}>_{2}|\frac {1}{2}>_{3} + |\frac
{1}{2}>_{2}|-\frac {1}{2}>_{3})]$  with zero spin even if neither
2 nor 3 is described by any spin eigen state. Finally phi meson as
subsystem 4 is simultaneously described by spin eigen state $|1>$
with spin 1. On the other hand, after DIS measurement, with
probability $\frac [{1}{3}\simeq 0.33$, measured quark, 1, is
described by spin eigen state $|\frac {1}{2}>_{1}$ with spin
$\frac {1}{2}$ equivalent to proton spin. Simultaneously, two
other quarks, 2 and 3, are described by spin eigen states $|-\frac
{1}{2}>_{2}$ and $|-\frac {1}{2}>_{3}$ with spins $-\frac {1}{2}$
and $-\frac {1}{2}$ both opposite to proton spin. Finally phi
meson as subsystem 4 is simulatenously again described by spin
eigen state $|1>$ with spin 1. Obviously, mentioned predictions
are in an excellent agreement with real DIS experiments.

Thus, suggested structure of the proton as a supersystem of two u
quarks, one d quark and one phi vector meson can effectively solve
proton spin crisis, but also it opens other interesting
possibility. Namely, as it is well-known, phi meson consists of
one strange quark $s$ and one anti strange quark $\bar {s}$. In
this way phi meson represents a system with almost totally
effectively hidden strange flavor. So, if phi meson must necessary
be introduced in the minimal proton structure necessary for a
correct description of the proton spin, it would mean that proton
is not an exactly nonstrange, but only almost totally effectively
hidden strange particle. It opens a new door for solution of the
old unsolved problem of the meaning of the quark and lepton
families. We paraphrase, "all lucky families are lucky at the unit
way".

Finally, we can extremely shortly consider supersymmetry, i.e.
SUSY crisis induced by definite absence (to this day) of the
positive detection of the SUSY predictions in many domains where
such detections would be occurred (WMAP, LHC, ATLAS, etc.). There
are many different attempts of the solutions of SUSY crisis. Some
of these solutions completely reject SUSY even if, as it is
well-known, SUSY solves many extremely hard conceptual problems in
the quantum field theory (GUT etc.). Other solutions suggest
necessity for the extension of the minimal supersymmetric standard
model (including or not concept of the spontaneous SUSY breaking),
however some analyses point out that such extensions are less
normal than initial MSSM.

We shall suggest a possible "first step" toward simple solution of
the SUSY crisis using the following superexclusion principle.

We suggest that all SUSY particles can be separated in two
disjunctive classes. In the first class there are so-called usual
SUSY particles (e.g. electron, photon, neutrino, Higgs boson,
graviton, … ) that can exist actually ("without interactions") or
virtually ("within interaction"). In the second class there are
supersymmetric, or shortly, s-partners of the usual SUSY particles
(e.g. selectron, photino, neutralino, Higgsino, gravitino, … ),
simply called unusual SUSY particles, that can exist
(super)exclusively virtually ("within interaction") but never
actually ("without interactions").

Suggested superexclusion principle simply explains why there is no
positive detection of the SUSY predictions (referring on the
characteristics of the unusual SUSY particles) on the one hand. On
the other hand suggested superexclusion principle admits use of
the SUSY for solution of the hard conceptual problems within
quantum field theories (e.g. GUT etc.) since it does nod forbid
diagrams with unusual SUSY particles.

Of course, here and now, we cannot yield any deeper explanation of
the  superexclusion principle (we hope that this concept
represents only a "deep truth" in the Niels Bohr sense of words).

In conclusion we can only repeat and point out the following.  In
this work we suggest a simple theoretical model of the proton able
to effectively solve proton spin crisis. Within domain of
applicability of this simple model proton consists only of two u
quarks and one d quarks (two of which have spin opposite to proton
and one identical to proton) and one neutral vector phi meson
(with spin two times larger than proton spin and directed
identically to proton spin). This model is in full agreement not
only with existing DIS experiments, but also with spin and
electric charge conservation as well as in a satisfactory
agreement with rest mass-energy conservation (since phi meson mass
is close to proton rest mass). Our model opens an interesting
possibility of the solution of the quarks and leptons families
problem (proton is not an absolutely non-strange particle, but
only a particle with almost totally effectively hidden strange).
Also we suggest a simple "first step" toward the solution of the
supersymmetry crisis using so-called superexclusion principle.
According to this principle usual particles (electron, neutrino, …
) can exist actually and virtually, while their supersymmetric
partners, s-particles (selectron, neutralino, …) can exist
(super)exclusively virtually but not actually.

\vspace{0.5cm}

{\large \bf References}

\begin{itemize}

\item [[1]] A. Ashman {\it et al}, Phys. Lett. B  {\bf 206} (1988) 394
\item [[2]] E. S. Ageev {\it et al}, Phys. Lett. B  {\bf 612} (2005) 154 ; hep-ex/0501073
\item [[3]] V. I. Alexakhin {\it et al}, Phys. Lett. B  {\bf 647} (2005) 8 ; hep-ex/0501073
\item [[4]] A. Airapetian {\it et al}, Phys. Rev. D {\bf 75} (2007 ) 012007 ;
hep-ex/0609.039
\item [[5]] J. Hansson, Progress in Physics {\bf 3} (2010) 52 ;  hep-ph/0304.225

\end {itemize}

\end {document}